\documentclass{phb-proc4-auth}
\usepackage{graphicx}
\usepackage{amssymb}

\begin{document}

\begin{frontmatter}
\journal{SCES '04}
\title{4-pole analysis of the two-dimensional Hubbard model}
\author{Satoru Odashima\corauthref{1}},
\author{Adolfo Avella},
\author{Ferdinando Mancini}
\address{Dipartimento di Fisica "E.R. Caianiello" - Unit\`{a} di Ricerca INFM di Salerno \\
Universit\`{a} degli Studi di Salerno, I-84081 Baronissi (SA),
Italy} \corauth[1]{Corresponding Author: Dipartimento di Fisica
"E.R. Caianiello", Universit\`{a} degli Studi di Salerno, Via S.
Allende, I-84081 Baronissi (SA), Italy.  Phone:  +39 089 965228.
Fax: +39 089 965275. E-mail: odashima@sa.infn.it}
\begin{abstract}
The electronic states of the two-dimensional Hubbard model are
investigated by means of a 4-pole approximation within the
Composite Operator Method. In addition to the conventional Hubbard
operators, we consider other two operators, which come from the
hierarchy of the equations of motion. These operators carry
information regarding surrounding spin and charge configurations.
By means of this operatorial basis, we can study the physics
related to the energy scale $J=4t^2/U$. Results present, in
addition to the main two bands, a quasi-particle peak at the Fermi
level, shadow bands and band flatness at ($\pi$, 0) point.
\end{abstract}
\end{frontmatter}

The main gross feature of the Hubbard model is the splitting of
the band with the formation of a gap of the order $U$. This
physics can be well understood in terms of the two Hubbard
operators, which mainly describe the two Hubbard sub-bands.
However, in order to catch low-energy features related to the
energy scale $J=4t^2/U$, it is necessary to take into account
higher-order operators, which carry information regarding
nearest-neighbor spin and charge correlation effects. Along this
line, we analyze the Hubbard model by means of a 4-pole
approximation within the Composite Operator Method
\cite{COM01,COM02,COM03,COM04}. In addition to the conventional
Hubbard operators, we consider other two operators that come from
the hierarchy of the equations of motion. The resulting band
structure will be much richer than the 2-pole one.

The two-dimensional Hubbard Hamiltonian reads as follows,
\begin{equation}
H = \sum_{\mathbf{ij}\sigma}(t_{\mathbf{ij}}
 -\mu\delta_{\mathbf{ij}})c^{\dagger}_{\sigma}(i)c_{\sigma}(j)
 +U\sum_{\mathbf{i}}n_{\uparrow}(i)n_{\downarrow}(i),
\end{equation}
where $c^{\dagger}_{\sigma}(i)$ and $c_{\sigma}(i)$ are creation
and annihilation operators of electrons with spin $\sigma$ at the
site $\mathbf{i}\ [ i=(\mathbf{i},t)]$, respectively. $n_{\sigma}(i) = c^{\dagger}_{\sigma}(i)
c_{\sigma}(i)$, $\mu$ is the chemical potential, $t_{\mathbf{ij}}=- 4 t
\alpha_{\mathbf{ij}}$, $\alpha[{\bf k}]=\mathcal{F} [\alpha_{\mathbf{ij}} ]
=\frac{1}{2} (\cos(k_x a)+\cos(k_y a))$, $a$ is the lattice
constant, $\mathcal{F}$ is the Fourier transform, $U$ is the
on-site Coulomb repulsion. We define the following operatorial
basis,
\begin{equation}
\psi_{A\sigma}(i)=\left( \begin{array}{l}
 \xi_{\sigma}(i) \\
 \eta_{\sigma}(i)
\end{array}\right).
\end{equation}
$\xi_{\sigma}(i)=c_{\sigma}(i)\left( 1-n_{-\sigma}(i) \right)$ and
$\eta_{\sigma}(i)=c_{\sigma}(i)n_{-\sigma}(i)$ describe the
transitions $n(i)=0 \leftrightarrow 1$ and $1 \leftrightarrow 2$,
respectively. The equations of motion of $\psi_{A\sigma}(i)$ give
\begin{equation}
\left\{ \begin{array}{l} \displaystyle i \frac{\partial}{\partial
t}\xi_{\sigma}(i) = -\mu \xi_{\sigma}(i)
 -4t\left[ c^{\alpha}_{\sigma}(i)+\pi_{\sigma}(i) \right] \vspace*{1mm}\\
\displaystyle i \frac{\partial}{\partial t}\eta_{\sigma}(i) =
(-\mu +U)\eta_{\sigma}(i)
 +4t\pi_{\sigma}(i).
\end{array} \right.
\end{equation}
where $\pi_{\sigma}(i)=-n_{-\sigma}(i)c^{\alpha}_{\sigma}(i)
 +c^{\dagger}_{-\sigma}(i)c_{\sigma}(i)c^{\alpha}_{-\sigma}(i)
 +c_{\sigma}(i)c^{\alpha\dagger}_{-\sigma}(i)c_{-\sigma}(i)$ with
$c^{\alpha}_{\sigma}(i)=\sum_{\bf j}\alpha_{\bf ij}c_{\sigma}({\bf
j},t)$.
Now, we divide $\pi_{\sigma}(i)$ into two operators
$\pi_{\sigma}(i)=\xi_{s\sigma}(i)+\eta_{s\sigma}(i)$ in the same
manner as we have done with
$c_{\sigma}(i)=\xi_{\sigma}(i)+\eta_{\sigma}(i)$. Then, we define
a new operator set
\begin{equation}
\psi_{B\sigma}(i)=\left( \begin{array}{l}
 \xi_{s\sigma}(i) \\
 \eta_{s\sigma}(i)
\end{array}\right)
\end{equation}
with
\begin{equation}
\begin{array}{l}
\xi_{s\sigma}(i)=-n_{-\sigma}(i)\xi^{\alpha}_{\sigma}(i)
 +c^{\dagger}_{-\sigma}(i)c_{\sigma}(i)\xi^{\alpha}_{-\sigma}(i) \\
 \hspace*{14mm} +c_{\sigma}(i)\eta^{\alpha\dagger}_{-\sigma}(i)c_{-\sigma}(i) \vspace*{1mm} \\
\eta_{s\sigma}(i)=-n_{-\sigma}(i)\eta^{\alpha}_{\sigma}(i)
 +c^{\dagger}_{-\sigma}(i)c_{\sigma}(i)\eta^{\alpha}_{-\sigma}(i) \\
 \hspace*{14mm} +c_{\sigma}(i)\xi^{\alpha\dagger}_{-\sigma}(i)c_{-\sigma}(i).
\end{array}
\end{equation}
It is worth mentioning that $\psi_{B\sigma}(i)$ describes two-site
composite excitations \cite{COM05} as they carry information of
surrounding spin and charge configurations, and are eigenoperators
of the interaction term as $\xi_{\sigma}(i)$ and
$\eta_{\sigma}(i)$.

In the present paper, we choose as operatorial basis
\begin{equation} \label{basis}
\psi_{\sigma}(i)=\left( \begin{array}{l}
 \psi_{A\sigma}(i) \\
 \psi_{B\sigma}(i)
\end{array}\right).
\end{equation}

Within the Composite Operator Method, once we choose a
$n$-component spinorial basis $\psi$, the equations of motion take the
general form
\begin{equation}
i\frac{\partial}{\partial t}\psi(i)=\sum_{\bf j}\epsilon({\bf
i},{\bf j}) \psi({\bf j},t)+\delta j(i)
\end{equation}
where $\epsilon({\bf k})=m({\bf k})I^{-1}({\bf k})$ is $n \times
n$ matrix as $I({\bf k})=\mathcal{F}\langle \{ \psi({\bf i},t),
\psi^{\dagger}({\bf j},t) \} \rangle$ and $m({\bf
k})=\mathcal{F}\langle \{ i\frac{\partial}{\partial t}\psi({\bf
i},t), \psi^{\dagger}({\bf j},t) \} \rangle$. If we neglect
$\delta j$ we obtain a pole structure for the Green's function
\begin{equation}
G(\omega, {\bf k})=\sum_{i=1}^n\frac{\sigma_{i}({\bf k})}{\omega
-E_{i}({\bf k})}.
\end{equation}
The 2-pole solution with $\psi=\psi_{A}$ have been discussed in
detail in Refs.~\cite{COM02,COM03}. In the present paper, we
analyze a 4-pole solution by means of the basis (\ref{basis}).
According to this, we have
\begin{equation}
I({\bf k})=\left(\begin{array}{cc}
  I_{AA}({\bf k}) & I_{AB}({\bf k}) \\
  I_{AB}({\bf k}) & I_{BB}({\bf k}) \\
\end{array}\right)
\end{equation}
\begin{equation}
\epsilon({\bf k})=\left(\begin{array}{cc}
  \epsilon_{AA}({\bf k}) & \epsilon_{AB}({\bf k}) \\
  \epsilon_{BA}({\bf k}) & \epsilon_{BB}({\bf k}) \\
\end{array}\right)
\end{equation}
The main difficulty of the present formulation is the evaluation
of $I_{BB}({\bf k})$ and $\epsilon_{BB}({\bf k})$. The other elements
can be simply computed either explicitly or in terms of those latter.
For $I_{BB}({\bf k})$, we calculate anticommutators explicitly and
decouple the higher-order correlation functions preserving
particle-hole symmetry and hermiticy. For $\epsilon_{BB}({\bf
k})$, we use the simplified equations of motion discussed in
Ref.~\cite{COM05},
\begin{equation}
\begin{array}{l}
i\frac{\partial}{\partial t}\xi_{s\sigma}(i) \simeq
-\mu\xi_{s\sigma}(i) +4t \left[ \frac{1}{2}\eta_{\sigma}(i)
+\xi^{\alpha}_{s\sigma}(i)
+2\eta^{\alpha}_{s\sigma}(i) \right] \vspace*{1mm} \\
i\frac{\partial}{\partial t}\eta_{s\sigma}(i) \simeq
(-\mu+U)\eta_{s\sigma}(i) +4t \left[
\frac{1}{4}\eta_{\sigma}(i)+\xi^{\alpha}_{s\sigma}(i) \right],
\end{array}
\end{equation}
where we neglect terms that give diffusion processes over three
sites. This procedure simply gives $\epsilon_{BB}({\bf k})$ by
inspection. The correlation functions appearing in the $I$ and
$\epsilon$ are self-consistently determined by means of Green's
function and the local algebra constraints \cite{COM04}.

In Fig.~1, we present the density of states and the dispersion
relation. It is worth noticing that the above formulation is
applicable in any dimension. In addition to the main Hubbard band
structure, our results show a coherent peak around the Fermi level
and shadow bands originated by the antiferromagnetic correlations.
The details of formulae and a more extended analysis will be
presented elsewhere.

\begin{figure}
\begin{center}
\includegraphics[width=6.3cm]{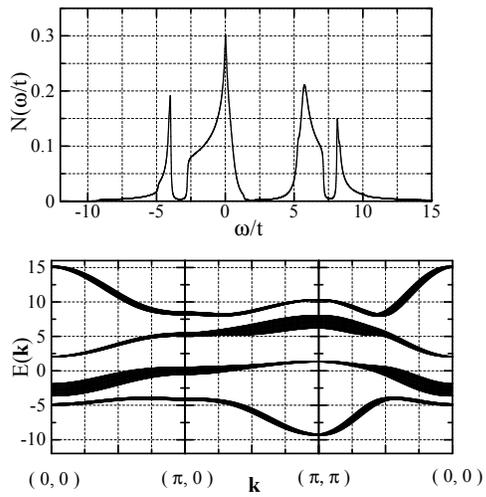}
\end{center}
\caption{The density of states (top) and the dispersion relation
(bottom) for $U=8t$, $n=0.9$ and $T=0.01t$. The width of
dispersion line gives the intensity of peak.}
\end{figure}

\end{document}